# Data Science as Political Action: Grounding Data Science in a Politics of Justice

Ben Green*

**Abstract:** In response to public scrutiny of data-driven algorithms, the field of data science has adopted ethics training and principles. Although ethics can help data scientists reflect on certain normative aspects of their work, such efforts are ill-equipped to generate a data science that avoids social harms and promotes social justice. In this article, I argue that data science must embrace a political orientation. Data scientists must recognize themselves as political actors engaged in normative constructions of society and evaluate their work according to its downstream impacts on people's lives. I first articulate why data scientists must recognize themselves as political actors. In this section, I respond to three arguments that data scientists commonly invoke when challenged to take political positions regarding their work. In confronting these arguments, I describe why attempting to remain apolitical is itself a political stance—a fundamentally conservative one—and why data science's attempts to promote "social good" dangerously rely on unarticulated and incrementalist political assumptions. I then propose a framework for how data science can evolve toward a deliberative and rigorous politics of social justice. I conceptualize the process of developing a politically engaged data science as a sequence of four stages. Pursuing these new approaches will empower data scientists with new methods for thoughtfully and rigorously contributing to social justice.

## 1 Introduction

The field of data science has entered a period of reflection and reevaluation.① Alongside its rapid growth in both size and stature in recent years, data science has become beset by controversies and scrutiny. Machine learning algorithms that guide decisions in areas such as hiring, healthcare, criminal sentencing, and welfare are often biased, inscrutable, and proprietary[1−6]. Algorithms that drive social media feeds manipulate people's emotions[7], spread misinformation[8], and amplify political extremism[9]. Facilitating these and other algorithms are massive datasets, often gained illicitly or without meaningful consent, that reveal sensitive and intimate information about people[10−13].

Many individuals and organizations responded to these controversies by advocating for a focus on ethics in computing training and practice[14]. Universities have created new courses that train students to consider the ethical implications of computer science[15−18]; one crowdsourced list includes more than 300 such classes[19]. Former US Chief Data Scientist D. J. Patil has argued that data scientists need a code of ethics[20]. The Association for Computing Machinery (ACM), the world's largest educational and scientific computing society, updated its Code of Ethics and Professional Conduct in 2018 for the first time since 1992[21]. The broad motivation behind these efforts is the assumption that, if only data scientists were more attuned to the

• Ben Green is with the Society of Fellows and the Gerald R. Ford School of Public Policy, University of Michigan, Ann Arbor, MI 48109, USA. E-mail: bzgreen@umich.edu.
∗ To whom correspondence should be addressed.
Manuscript received: 2021-05-20; revised: 2021-11-19; accepted: 2021-11-25

① Throughout this article, "data science" encompasses the use of computational methods (including artificial intelligence and machine learning) to derive patterns from data in order to make predictions about the future. In this sense, a data scientist is anyone who works with data and algorithms in these settings. My particular focus is on the application of data science methods to social and political contexts.

ethical implications of their work, many harms associated with data science could be avoided[14].

Although emphasizing ethics is an important step in data science's development toward greater socially responsibility, it is an insufficient response to the broad issues of social justice that are implicated by data science.② As described in the introductory article for this special issue, technology ethics as applied in practice suffers from four significant limitations[14]. First, technology ethics principles are abstract and lack mechanisms to ensure that engineers follow ethical principles. Second, technology ethics has a myopic focus on individual engineers and on technology design, overlooking the structural sources of technological harms. Third, technology ethics is subsumed into corporate logics and practices rather than substantively altering behavior. All told, the rise of technology ethics often reflects a practice dubbed "ethics-washing": tech companies deploying the language of ethics to resist more structural reforms that would curb their power and profits.

Thus, while ethics provides useful frameworks to help data scientists reflect on their practice and the impacts of their work, these approaches are insufficient for generating a data science that avoids social harms and that promotes social justice. The normative responsibilities of data scientists cannot be managed through to a narrow professional ethics that lacks normative weight and supposes that, with some reflection and a commitment to best practices, data scientists will make the "right" decisions that lead to "good" technology. Instead of relying on vague moral principles that obscure the structural drivers of injustice, data scientists must engage in politics: the process of negotiating between competing perspectives, values, and goals.

In other words, we must recognize data science as a form of political action. Data scientists must recognize themselves as political actors engaged in normative constructions of society. In turn, data scientists must evaluate their efforts according to the downstream impacts on people's lives.

By politics and political, I do not refer directly to partisan or electoral debates about specific parties and candidates. Instead, I invoke these terms in a broader sense that transcends activity directly pertaining to the government, its laws, and its representatives. Two aspects of politics are paramount. First, politics is everywhere in the social world. As defined by politics professor Adrian Leftwich, "politics is at the heart of all collective social activity, formal and informal, public and private, in all human groups, institutions, and societies"[23]. Second, politics has a broad reach. Political scientist Harold Lasswell describes politics as "who gets what, when, how"[24]. The "what" here could mean many things: money, goods, status, influence, respect, rights, and so on. Understood in these terms, politics comprises any activities that affect or make claims about the who, what, when, and how in social groups, both small and large.

Data scientists are political actors in that they play an increasingly powerful role in determining the distribution of rights, status, and goods across many social contexts. As data scientists develop tools that inform important social and political decisions—who receives a job offer, what news people see, where police patrols—they shape social outcomes around the world. Data scientists are some of today's most powerful (and obscured) political actors, structuring how institutions conceive of problems and make decisions.

This article will justify and develop the notion of data science as political action. My argument raises two questions: (1) Why must data scientists recognize themselves as political actors? and (2) How can data scientists ground their practice in a politics of social justice? The two primary sections of this article will take up these questions in turn.

My aim is to support data science toward playing a more productive role in promoting equity and social justice. I do not intend to stop data science in its tracks, critique individual practitioners, or discourage data scientists from working on social problems. The path ahead does not require data scientists to abandon their technical expertise, but it does require data scientists to expand their notions of what problems to work on and how to engage with society. This process may involve an uncomfortable period of change. But I am confident that exciting new areas for research and practice will emerge, producing a field that can contribute to a more egalitarian and just society.

② In *Black Feminist Thought*, Patricia Hill Collins defines a "social justice project" as "an organized, long-term effort to eliminate oppression and empower individuals and groups within a just society". Oppression, she writes, is "an unjust situation where, systematically and over a long period of time, one group denies another group access to the resources of society"[22].

## 2 Why Must Data Scientists Recognize Themselves as Political Actors?

The first part of this article will attempt to answer this question in the form of a dialogue with a well-intentioned skeptic. I will respond to three arguments that are commonly invoked by data scientists when they are challenged to take political stances regarding their work. These arguments have been expressed in a variety of public and private settings and will be familiar to anyone who has engaged in discussions about the social responsibilities of data scientists.

These are by no means the only arguments proffered in this larger debate, nor do they represent any sort of unified position among data scientists. In practice, computer scientists are "diverse and ambivalent characters"[25] who engage in "nuanced, contextualized, and reflexive practices"[26]. Some computer science subfields (such as CSCW[27]) have long histories of engaging with sociotechnical practices and normative implications, while others (such as the ACM Conference on Fairness, Accountability, and Transparency (FAccT)) are actively developing such approaches. Nonetheless, in my experience, the three positions considered here are the most common and compelling arguments made against a politically oriented data science. Any promotion of a more politically engaged data science must contend with them.

### 2.1 Argument 1: "I am just an engineer"

This first argument represents a common attitude among engineers. In this view, although engineers develop new tools, their work does not determine how a tool will be used. Artifacts are seen as neutral objects that lack any inherent normative character and that can simply be used in good or bad ways. By this logic, engineers bear no responsibility for the impacts of their creations.

It is common for data scientists to argue that the impacts of technology are unknowable. As one computer scientist who faced criticism for developing facial recognition software argued in defense of his work, "Anything can be used for good. Anything can be used for bad"[28]. Similarly, during a 2019 NeurIPS workshop, in which two panelists highlighted the harmful impacts of AI on communities of color, several computer scientists in the audience countered that it is impossible to know what the impacts of research will be or to prevent others from misusing products[29].

By articulating their limited role as neutral researchers, data scientists provide themselves with an excuse to abdicate responsibility for the social and political impacts of their work. When a paper that used neural networks to classify crimes as gang-related was challenged for its potentially harmful effects on minority communities, a senior author on the paper deflected responsibility by arguing, "It's basic research"[30].

Although it is common for engineers to see themselves as separate from politics, many scholars have thoroughly articulated how technology embeds politics and shapes social outcomes. As political theorist Langdon Winner describes, "technological innovations are similar to legislative acts or political foundings that establish a framework for public order that will endure over many generations. For that reason, the same careful attention one would give to the rules, roles, and relationships of politics must also be given to such things as the building of highways, the creation of television networks, and the tailoring of seemingly insignificant features on new machines. The issues that divide or unite people in society are settled not only in the institutions and practices of politics proper, but also, and less obviously, in tangible arrangements of steel and concrete, wires and semiconductors, and nuts and bolts"[31].

Even though technology does not conform to conventional notions of politics, it often shapes society in much the same way as laws, elections, and judicial opinions. In this sense, "the scientific workplace functions as a key site for the production of social and political order"[32]. Thus, as with many other types of scientists, data scientists possess "a source of fresh power that escapes the routine and easy definition of a stated political power"[33].

There are many examples of engineers developing and deploying technologies that, by structuring behavior and shifting power, shape aspects of society. As one example, Winner famously (and controversially[34, 35]) describes how Robert Moses designed the bridges over the parkways on Long Island, New York with low overpasses[31]. Moses purportedly did this to prevent buses (which predominantly carried lower-class and non-white urban residents) from navigating these parkways and accessing the parks to which they led.

Another historical example similarly demonstrates how the design of traffic technologies can have social and political ramifications. As historian Peter Norton describes, when automobiles were introduced onto city streets in the 1920s, they created chaos and conflict in the existing social order[36]. Many cities turned to traffic engineers as "disinterested experts" whose scientific methods could provide a neutral and optimal solution. But the engineers' solution contained unexamined assumptions and values, namely, that "traffic efficiency worked for the benefit of all". As traffic engineers changed the timings of traffic signals to enable cars to flow freely, their so-called solution "helped to redefine streets as motor thoroughfares where pedestrians did not belong". These actions by traffic engineers helped shape the next several decades of automobile-focused urban development in US cities.

Although these particular outcomes could be chalked up to unthoughtful design, any decisions that the traffic engineers made would have had some such impact: determining how to time streetlights requires judgments about what outcomes and whose interests to prioritize. Whatever they and the public may have believed, traffic engineers were never "just" engineers optimizing society "for the benefit of all". Instead, they were engaged in the process—via formulas and signal timings—of defining which street uses should be supported and which should be constrained. The traffic engineers may not have decreed by law that streets were for cars, but their technological intervention assured this outcome by other means.

Data scientists today risk repeating this pattern of designing tools with inherently political characters yet largely overlooking their own agency and responsibility. By imagining an artificially limited role for themselves, engineers create an environment of scientific development that requires few moral or political responsibilities. But this conception of engineering has always been a mirage. Developing any technology contributes to the particular "social contract implied by building that technological system in a particular form"[31].

Of course, we must also resist placing too much responsibility on data scientists. The point is not that, if only they recognized their social impacts, engineers could themselves solve social issues. Technology is at best just one tool among many for addressing complex social problems[37]. Nor should we uncritically accept the social influence that data scientists have. Having unelected and unaccountable technical experts make core decisions about governance away from the public eye imperils essential notions of how a democratic society ought to function. As Science, Technology, and Society (STS) scholar Sheila Jasanoff argues, "The very meaning of democracy increasingly hinges on negotiating the limits of the expert's power in relation to that of the publics served by technology"[38].

Nonetheless, the design and implementation of technology does rely, at some level, on trained practitioners. This raises several questions that animate the rest of this article. What responsibilities should data scientists bear? How must data scientists reconceptualize their scientific and societal roles in light of these responsibilities?

### 2.2 Argument 2: "Our job is not to take political stances"

Data scientists adhering to this second argument likely accept the response to Argument 1 but feel stuck, unsure how to appropriately act as more than "just" an engineer. "Sure, I am developing tools that impact people's lives", they may acknowledge, before asking, "But is not the best thing to just be as neutral as possible?"

Although it is understandable how data scientists come to this position, their desire for neutrality suffers from two important failings. First, neutrality is an unachievable goal, as it is impossible to engage in science or politics without being influenced by one's background, values, and interests. Second, striving to be neutral is not itself a politically neutral position. Instead, it is a fundamentally conservative one.③

An ethos of objectivity has long been prevalent among scientists. Since the nineteenth century, objectivity has evolved into a set of widespread ethical and normative scientific practices. Conducting good science—and being a good scientist—meant suppressing one's own perspective so that it would not contaminate the interpretations of observations[39].

Yet this conception of science was always rife with contradictions and oversights. Knowledge is shaped and bounded by the social contexts that generated it. This insight forms the backbone of standpoint theory, which articulates that "nothing in science can be protected from cultural influence—not its methods, its research technologies, its conceptions of nature's fundamental ordering principles, its other concepts, metaphors,

③ I use conservative here in the sense of maintaining the status quo rather than in relation to any specific political party or movement.

models, narrative structures, or even formal languages"[40]. Although scientific standards of objectivity account for certain kinds of individual subjectivity, they are too narrowly construed: "methods for maximizing objectivism have no way of detecting values, interests, discursive resources, and ways of organizing the production of knowledge that first constitute scientific problems, and then select central concepts, hypotheses to be tested, and research designs"[40].

These processes make the supposedly objective scientific "gaze from nowhere" nothing more than "an illusion"[41]. Every aspect of science is imbued with the characteristics and interests of those who produce it. This does not invalidate every scientific finding as arbitrary, but points to science's contingency and reliance on its practitioners: all research and engineering are developed within particular institutions and cultures and with particular problems and purposes in mind.

Just as it is impossible to conduct science in any truly neutral way, there is no such thing as a neutral (or apolitical) approach to politics. As philosopher Roberto Unger writes, political neutrality is an "illusory and ultimately idolatrous goal" because "no set of practices and institutions can be neutral among conceptions of the good"[42].

Instead of being neutral and apolitical, attempts to be neutral and apolitical embody an implicitly conservative politics. Because neutrality does not mean value-free—it means acquiescence to dominant social and political values, freezing the status quo in place. Neutrality may appear to be apolitical, but that is only because the status quo is taken as a neutral default. Anything that challenges the status quo—which efforts to promote social justice must do by definition—will therefore be seen as political. But efforts for reform are no more political than efforts to resist reform or even the choice simply to not act, both of which preserve existing systems.

Although surely not the intent of every scientist and engineer who strives for neutrality, broad cultural conceptions of science as neutral entrench the perspectives of dominant social groups, who are the only ones entitled to legitimate claims of neutrality. For example, many scholars have noted that neutrality is defined by a masculine perspective, making it impossible for women to be seen as objective or for neutral positions to consider female standpoints[40, 43–45]. The voices of Black women are particularly subjugated as partisan and anecdotal[22]. Because of these perceptions, when people from marginalized groups critique scientific findings, they are cast off as irrational, political, and representing a particular perspective[41]. In contrast, the practices of science and the perspectives of the dominant groups that uphold it are rarely considered to suffer from the same maladies.

Data science exists on this political landscape. Whether articulated by their developers or not, machine learning systems already embed political stances. Overlooking this reality merely allows these political judgments to pass without scrutiny, in turn granting data science systems with more credence and legitimacy than they deserve.

Predictive policing algorithms offer a particularly pointed example of how striving to remain neutral entrenches and legitimize existing political conditions. The issue is not simply that the training data behind predictive policing algorithms are biased due to a history of overenforcement in minority neighborhoods. In addition, our very definitions of crime and how to address it are the product of racist and classist historical processes. Dating back to the eras of slavery and reconstruction, cultural associations of Black men with criminality have justified extensive police forces with broad powers[46]. The War on Drugs, often identified as a significant cause of mass incarceration, emerged out of an explicit agenda by the Nixon administration to target people of color[47].④ Meanwhile, crimes like wage theft are systemically underenforced by police and do not even register as relevant to conversations about predictive policing.⑤

Moreover, predictive policing rests on a model of policing that is itself unjust. Predictive policing software could exist only in a society that deploys vast punitive resources to prevent social disorder, following "broken

④ As Nixon's special counsel John Ehrlichman explained years later, "We knew we could not make it illegal to be either against the war or black. But by getting the public to associate the hippies with marijuana and blacks with heroin, and then criminalizing both heavily, we could disrupt those communities. We could arrest their leaders, raid their homes, break up their meetings, and vilify them night after night on the evening news. Did we know we were lying about the drugs? Of course we did."[48]

⑤ Wage theft occurs when employers deny their employees the wages or benefits to which they are legally entitled (e.g., not paying employees for overtime work). Wage theft steals more value than all other kinds of theft (such as burglaries) combined, typically carried out by business owners against low-income workers[49].

windows" tactics. Policing has always been far from neutral: "the basic nature of the law and the police, since its earliest origins, is to be a tool for managing inequality and maintaining the status quo"[50]. The issues with policing are not flaws of training or methods or "bad apple" officers, but are endemic to policing itself[46, 50].

Against this backdrop, choosing to develop predictive policing algorithms is not neutral. Accepting common definitions of crime and how to address it may seem to allow data scientists to remove themselves from politics, but instead upholds historical politics of social hierarchy.

Although predictive policing represents a notably salient example of how data science cannot be neutral, the same could be said of all applied data science. Biased data are certainly one piece of the story, but so are existing social and political conditions, definitions and classifications of social problems, and the set of institutions that respond to those problems. None of these factors are neutral and removed from politics. And while data scientists are of course not responsible for creating these aspects of society, they are responsible for choosing how to interact with them. Neutrality in the face of injustice only reinforces that injustice. When engaging with aspects of the world steeped in history and politics, in other words, it is impossible for data scientists to not take political stances.

I do not mean to suggest that every data scientist should share a singular political vision—that would be wildly unrealistic. It is precisely because the field (and world) hosts a diversity of normative perspectives that we must surface political debates and recognize the role they play in shaping data science practice. Nor is my argument meant to suggest that articulating one's political commitments is a simple task. Normative ideals can be complex and conflicting, and one's own principles can evolve over time. Data scientists need not have precise answers about every political question. However, they must act in light of articulated principles and grapple with the uncertainty that surrounds these ideals.

### 2.3 Argument 3: "We should not let the perfect be the enemy of the good"

Following the responses to Arguments 1 and 2, data scientists asserting this third argument likely acknowledge that their creations will unavoidably have social impacts and that neutrality is not possible. Yet still holding out against a thorough political engagement, they fall back on a seemingly pragmatic position: because data science tools can improve society in incremental but important ways, we should support their development rather than argue about what a perfect solution might be.

Despite being the most sophisticated of the three arguments, this position suffers from several underdeveloped principles. First, data science lacks robust theories regarding what "perfect" and "good" actually entail. As a result, the field typically adopts a superficial approach to reform that involves making vague (almost tautological) claims about what social conditions are desirable. Second, this argument fails to articulate how to evaluate or navigate the relationship between the perfect and the good. Efforts to promote social good thus tend to take for granted that technology-centric incremental reform is an appropriate strategy for social progress. Yet, considered from a perspective of substantive equality and anti-oppression, many data science efforts to do good are not, in fact, consistently doing good.

#### 2.3.1 Data science lacks a thorough definition of "social good"

Across the broad world of data science, from academic institutes to conferences to companies to volunteer organizations, "social good" (or just "good") has become a popular term. Numerous universities across the United States and Europe have hosted the Data Science for Social Good Summer Fellowship.⑥ Several major computer science conferences have hosted AI for Social Good workshops,⑦ and in 2014 the theme of the entire ACM SIGKDD Conference on Knowledge Discovery and Data Mining (KDD) was "Data Mining for Social Good".⑧ Since 2014, the company Bloomberg has hosted an annual Data for Good Exchange.⑨ The non-profit Delta Analytics strives to promote "Data-driven solutions for social good".⑩

While this energy to do good among the data science community is both commendable and exciting, the field has not developed (nor even much debated) any working definitions of the term "social good" to guide its efforts. Instead, the field seems to operate on a "know it when

⑥ http://www.dssgfellowship.org
⑦ https: //aiforsocialgood.github.io/
⑧ https: //www.kdd.org/kdd2014/
⑨ https: //www.bloomberg.com/company/d4gx/
⑩ http://www.deltanalytics.org

you see it" approach, relying on rough proxies such as crime = bad, poverty = bad, and so on. The term's lack of precision prompted one of Delta Analytics' founders to write that "'data for good' has become an arbitrary term to the detriment of the goals of the movement"[51]. The notable exception is Mechanism Design for Social Good (MD4SG), which articulates a clear research agenda "to improve access to opportunity, especially for communities of individuals for whom opportunities have historically been limited"[52].

In fact, the term "social good" lacks a thorough definition even beyond the realm of data science. It is not defined in dictionaries like Merriam-Webster, the Oxford English Dictionary, and Dictionary.com, nor does it have a page on Wikipedia.⑪ To find a definition one must look to the financial education website Investopedia, which defines social good as "something that benefits the largest number of people in the largest possible way, such as clean air, clean water, healthcares, and literacy"[54]. There is, of course, extensive literature (spanning philosophy, STS, and other fields) that considers what is socially desirable, yet data science efforts to promote "social good" rarely reference this literature.

This lack of definition leads to "data science for social good" projects that span a wide range of conflicting political orientations. For example, some work under the "social good" umbrella is explicitly developed to enhance police accountability and promote non-punitive alternatives to incarceration[55, 56]. In contrast, other work under the "social good" label aims to enhance police operations. One such paper aimed to classify gang crimes in Los Angeles[30, 57]. This project involved taking for granted the legitimacy of the Los Angeles Police Department's gang data—a notoriously biased type of data[58] from a police department that has a long history of abusing minorities in the name of gang suppression[50]. That such politically disparate and conflicting work could be similarly characterized as "social good" should prompt a reconsideration of the core terms and principles. When the term encompasses everything, it means nothing.

The point is not that there exists a single optimal definition of "social good", nor that every data scientist should agree on one set of principles. Instead, there is a multiplicity of perspectives that must be openly acknowledged to surface debates about what "good" actually entails. Currently, however, the field lacks the language and perspective to sufficiently evaluate and debate differing visions of what is "good". By framing their notions of "good" in such vague and undefined terms, data scientists get to have their cake and eat it too: they can receive praise and publications based on broad claims about solving social challenges, while avoiding substantive engagement with social and political impacts.

Most dangerously, data science's vague framing of social good allows those already in power to present their normative judgments about what is "good" as neutral facts that are difficult to challenge. As discussed in Section 2.2, neutrality is an impossible goal and attempts to be neutral tend to reinforce the status quo. Thus, if the field does not openly debate definitions of "perfect" and "good", the assumptions and values of dominant groups will tend to win out. Projects that purport to enhance social good but fail to reflexively engage with the political context are likely to reproduce the exact forms of social oppression that many working towards "social good" seek to dismantle.⑫

### 2.3.2 Pursuing an incremental "good" can reinforce oppression

Even if data scientists acknowledge that "social good" is often poorly defined, they may still adhere to the argument that "we should not let the perfect be the enemy of the good". "After all", they might say, "is not some solution, however imperfect, better than nothing?" As one paper asserts, "we should not delay solutions over concerns of optimal" outcomes[60].

At this point the second failure of Argument 3 becomes clear: it tells us nothing about the relationship between the perfect and the good. Data science has thus far not developed any rigorous methodology for considering the relationship between algorithmic interventions and social impacts. Although data scientists generally acknowledge that data science cannot provide perfect solutions to social problems, the field typically takes for granted that incremental reforms using data science contribute to the "social good". On this logic, we should applaud any attempts to alleviate issues such as crime, poverty, and discrimination. Meanwhile, because "the perfect" represents an

⑪ Searching Wikipedia for "social good" automatically redirects to the page for "common good", a term similarly undefined in data science parlance[53].

⑫ Reflexivity refers to the practice of treating one's own scientific inquiry as a subject of analysis[59].

unrealizable utopia we should not waste time and energy debating the ideal solution.

Although efforts to promote “social good” using data science can be productive,⑬ pursuing such applications without a rigorous theory of social change can lead to harmful consequences. A reform that seems desirable from a narrow perspective focused on immediate improvements can be undesirable from a broader perspective focused on long-term, structural reforms. Understood in these terms, the dichotomy between the idealized “perfect” and the incremental “good” is a false one: articulating visions of an ideal society is an essential step for developing and evaluating incremental reforms. In order to rigorously conceive of and compare potential incremental reforms, we must first debate and refine our conceptions of the society we want to create; following those ideals, we can then evaluate whether potential incremental reforms push society in the desired direction. Because there is a multiplicity of imagined “perfects”, which in turn suggest an even larger multiplicity of incremental “goods”, reforms must be evaluated based on what type of society they promote in both the short and long term. In other words, rather than treating any incremental reform as desirable, data scientists must recognize that different incremental reforms can push society down drastically different paths.

When attempting to achieve reform, an essential task is to evaluate the relationship between incremental changes and long-term agendas for a more just society. As social philosopher André Gorz proposes, we must distinguish between “reformist reforms” and “non-reformist reforms”[61]. Gorz explains, “A reformist reform is one which subordinates its objectives to the criteria of rationality and practicability of a given system and policy.” In contrast, a non-reformist reform “is conceived not in terms of what is possible within the framework of a given system and administration, but in view of what should be made possible in terms of human needs and demands”.

Reformist and non-reformist reforms are both categories of incremental reform, but they are conceived through distinct processes. Reformist reformers start within existing systems, looking for ways to improve them. In contrast, non-reformist reformers start beyond existing systems, looking for ways to achieve emancipatory social conditions. Because of the distinct ways that these two types of reforms are conceived, the pursuit of one versus the other can lead to widely divergent social and political outcomes.

The solutions proposed by data scientists are almost entirely reformist reforms. The standard logic of data science—grounded in accuracy and efficiency—tends toward accepting and working within the parameters of existing systems. Data science interventions are therefore typically proposed to improve the performance of a system rather than to substantively alter it. And while these types of reforms have value under certain conditions, such an ethos of reformist reforms is unequipped to identify and pursue the larger changes that are necessary across many institutions. This approach may even serve to entrench and legitimize the status quo. From the standpoint of existing systems, it is impossible to imagine alternative ways of structuring society—when reform is conceived in this way, “only the most narrow parameters of change are possible and allowable”[62].

In this sense, data science’s dominant strategy of pursuing a reformist, incremental good resembles a greedy algorithm: at every point in time, the strategy is to make immediate improvements in the local vicinity of the status quo. Although a greedy strategy can be useful for simple problems, it is unreliable in complex search spaces: we may quickly find a local maximum but will never reach a further-afield terrain of far better solutions. Moves that are immediately beneficial can be counterproductive for finding the global optimum. Similarly, although reformist reforms can lead to certain improvements, a strategy limited to reformist reforms cannot guide robust responses to complex political problems. Reforms that appear desirable within the narrow scope of a reformist strategy can be counterproductive for achieving structural reforms. Even though the optimal political solution is rarely achievable (and is often subject to significant debate), it is necessary to fully characterize the space of possible reforms and to evaluate how reliably different approaches can generate more egalitarian outcomes.

The US criminal justice system, a domain where data scientists are increasingly striving to do good, exemplifies the limits of a reformist mindset. Because criminal justice reform can be “superficial and deceptive”[63], it is necessary to couch reform efforts

⑬ See e.g., the set of projects completed by the Data Science for Social Good Fellowship: http://www.dssgfellowship.org/projects/.

within a broader vision of long-term, non-reformist change. This is the approach taken by the movement for police and prison abolition. Notably, prison abolitionists object to reforms that "render criminal law administration more humane, but fail to substitute alternative institutions or approaches to realize social order maintenance goals"[64]. Instead, abolitionists pursue only reforms that reduce or replace carceral responses to social disorder.

In contrast with this abolitionist ethos, most data science efforts to contribute "good" are grounded in the existing practices of the criminal justice system. A notable example is pretrial risk assessments. Even if they lead to incremental improvements, these tools legitimize policies that drive racial injustice and mass incarceration[65]. Meanwhile, an entirely separate incremental reform—an abolitionist and non-reformist (and non-technological) one—is possible: ending cash bail and pretrial detention. Recent surveys show public support for such reforms[66, 67].

Adopting pretrial risk assessments and abolishing pretrial detention appear to respond to the same problems, suggesting that these two reforms are aligned. However, these reforms derive from conflicting visions of the "perfect". Reformers supporting risk assessments accept pretrial detention as part of criminal justice system, aiming merely to improve the means by which people are selected for pretrial detention. Meanwhile, reformers aiming to abolishing pretrial detention reject pretrial detention, aiming to abolish the practice altogether. In other words, the debate about risk assessments hinges on political questions about how the criminal justice system should be structured. It is only by articulating our imagined perfects that we can even recognize the underlying tension between these two incremental reforms, let alone properly debate which one to pursue.

The point is not that data science is incapable of improving society. However, data science interventions must be evaluated against alternative reforms as just one of many options, rather than compared merely against the status quo as the only possible reform. There should not a default presumption that machine learning provides an appropriate reform for every problem.

In sum, attempts by data scientists to avoid politics overlook technology's social impacts, privilege the status quo, and narrow the range of possible reforms. The field of data science will be unable to meaningfully advance social justice without accepting itself as political. The question that remains is how it can do so.

## 3 How Can Data Scientists Ground Their Practice in Politics?

The first part of this article argued that data scientists must recognize themselves as political actors. Yet several questions remain: What would it look like for data science to be explicitly grounded in a politics of social justice? How might the field evolve toward this end?

I conceptualize the process of incorporating politics into data science as following four stages, with reforms at both the individual and the institutional/cultural levels. Stage 1 (Interest) involves data scientists becoming interested in working directly on addressing social issues. In Stage 2 (Reflection), the data scientists involved in that work come to recognize the politics that underlie these issues and their attempts to address them.⑭ This leads to Stage 3 (Applications), in which data scientists direct the methods at their disposal toward new problems. Finally, Stage 4 (Practice) involves the long-term project of developing new methods and structures that orient data science around a politics of social justice.

I discuss each stage in more detail below. While not every person or project will follow this precise trajectory, it presents a possible path for data scientists to incorporate politics into their practice. In fact, many data scientists already are following some version of these stages toward a politically informed data science.

### 3.1 Stage 1: Interest

The first step toward infusing a deliberate politics into data science is for data scientists to orient their work around addressing social issues. Such efforts are already well underway, from "data for good" programs to civic technology groups to the growing numbers of data scientists working in governments and non-profits. Although they may not have an articulated vision of "social good", many data scientists are eager to apply

⑭ Some might argue that the order of Stages 1 and 2 should be reversed: data scientists should reflect first, then act to address social issues. This would be the most responsible approach and is the practice that data scientists should follow in the long term. In my experience, however, data scientists' engagements with politics tend to begin with an interest in addressing social challenges, which then leads to reflection on the politics of data science. New pedagogical approaches could merge these two stages. For instance, a "public interest tech" program could integrate reflection on the political nature of data science into its efforts to apply data science in practice.

their work to pressing societal challenges.

However, relative to the excitement around such work, there is a dearth of opportunities for data scientists to apply their skills to an articulated vision of social benefit. Many academic departments and conferences tend not to consider such work to be valid research, companies can find more profit elsewhere, and governments and non-profits have few internal data science roles. Thus, many data scientists who want to do socially impactful work often settle for more traditional research or jobs, in which technical contributions and profit provide the primary imperatives.

Data science programs should work towards a model of "public interest technology" that trains data scientists to address social issues. This involves not simply adopting this label, but also providing methods, pathways, and a broader culture of support for data scientists to improve society. For example, data science programs should develop clinics where students provide technical and policy assistance to "clients" such as activists and government agencies. Programs should also provide funding and guidance for students to find internships and jobs focused on social impact.⑮

It is essential that "social good" and "public interest tech" programs prioritize social and political reforms over deploying technology. The driving goal should be to positively impact society rather than to develop sophisticated tools. This requires an attitude of agnosticism: "approaching algorithms instrumentally, recognizing them as just one type of intervention, one that cannot provide the solution to every problem"[68]. The more that data scientists work directly with governments, communities, and service providers (rather than on abstract technology problems), the more thoroughly they will come to see technology as an imperfect means rather than as an end in itself. Without this technology-agnostic focus on social impacts, efforts to apply data science to social problems will reproduce the issues described in Section 2.3 and will prevent progression to the following stages.

## 3.2 Stage 2: Reflection

As they work on data science for social good projects, data scientists will encounter the political nature of both the issues at hand and their own efforts to address these issues. To the extent that they maintain an open-minded and critical approach grounded in impact, data scientists will begin to reflect on political questions.

We have seen this process play out most clearly with respect to algorithmic bias and fairness. Where just a few years ago it was common to hear claims that data represents "facts" and that algorithms are "objective"[69, 70], today it is widely acknowledged within data science that data contains biases and that algorithms can discriminate. In addition to the annual ACM Conference on Fairness, Accountability, and Transparency (FAccT), there have been numerous workshops dedicated to these issues at major computer science conferences[71]. Moreover, there is also an emerging literature that articulates the limitations and politics of common approaches to studying and promoting algorithmic fairness[72–74].

Over time, data scientists must expand this critical and reflexive lens to increasingly interrogate how all aspects of their work are political. For example, returning to the discussion of predictive policing from Section 2.2, it is not sufficient to develop algorithms just with a recognition that crime data are biased. It is necessary to also recognize that our definitions of crime, the set of institutions that are tasked with responding to it, and the interventions that those institutions provide are all the result of historical political processes laden with discrimination.

Reflection of this sort is propelled by approaching research with an open mind and honoring the expertise of other disciplines, policymakers, and affected communities. Such reflection will be particularly enhanced by fluency in fields such as STS and critical algorithm studies. Exposure to these fields should become central to data science training programs, particularly those with an emphasis on applications of data science for social good. For data scientists hoping to improve society, familiarity with STS and related fields is just as essential as knowledge of databases and statistics.

## 3.3 Stage 3: Applications

In the short term, the insights provided in Stage 2 are not likely to shake the fundamental structures and practices of data science. Instead, these insights will empower data scientists to seek new applications for how existing data science methods can address injustice and shift power.

⑮ See e.g., a list of job boards and other resources that I have compiled: https: //www.benzevgreen.com/jobs/.

These effects will demonstrate how incorporating a political perspective into data science produces new directions for research and applications rather than a dead end.

Several frameworks can guide data scientists in these efforts. For example, André Gorz's schema of non-reformist reforms and the framework of prison abolition provide conceptual tools for moving beyond the false dichotomy between incremental and radical reform[61, 64]. The notion of "critical design" embodies a similar approach: in contrast to "affirmative design, which "reinforces how things are now", "critical design provides a critique of how things are now through designs that embody alternative social, cultural, technical, or economic values"[75]. A related framework is "anti-oppressive design", which provides "a guide for how best to expend resources, be it the choice of a research topic, the focus of a new social enterprise, or the selection of clients and projects, rather than relying on vague intentions or received wisdom about what constitutes good"[76].

At each stage of the research and design process, data scientists should evaluate their efforts according to these frameworks: Should the design of this algorithm be affirmative or critical? Would the implementation of this model represent a reformist or non-reformist reform? Would empowering our project partner with this system challenge or entrench oppression? Such analyses can help data scientists interrogate their notions of "good" to engage in non-reformist, critical, and anti-oppressive data science. These approaches can also help data scientists recognize situations in which non-technological reforms are more desirable than technological ones[37, 77].

This ethos of pursuing different, politically motivated data science applications can inform work in areas such as policing. One dimension of this shift involves a critical and anti-oppressive approach to selecting project partners. For example, some researchers explicitly articulate an intention to work with community groups and social service providers rather than with law enforcement, recognizing that the latter tend to contribute to structural oppression[55, 78, 79]. Another dimension of this shift involves orienting the analytic gaze away from individuals and towards institutions. One example of this work used machine learning to predict which police officers will be involved in adverse events such as racial profiling or inappropriate use of force[56]. Others have used new algorithmic methods to find evidence of racial bias in police behavior[80, 81].

Although Stage 3 represents a significant evolution of data science toward politics, it suffers from three notable shortcomings. First, it is possible to operate in Stage 3 without ever articulating an explicit politics. Although not raising a project's political motivations may enable some projects to pass without scrutiny, it does little to provide language or direction for other data scientists. The field will not evolve if political debates remain shrouded. Moreover, only relatively minor reforms could be successfully promoted in this covert manner: more significant reforms will likely be challenged and will advance only if they can be explicitly defended.

Second, existing data science methods have a limited ability to promote social justice. Because of data science's adherence to mathematical formalism, current methods are incapable of rigorously representing and reasoning about social contexts and political impacts[68]. Thus, even well-intentioned and seemingly well-designed data science tools can promote injustice[74].

Third, merely directing data science toward new applications remains fundamentally undemocratic: it allows data scientists to shape society without deliberation or accountability. In this frame, a cadre of data scientists—no matter their intentions or actions—retain an outsized power to shape institutions and decision-making processes. Even when their actions are grounded in anti-oppressive ideals, the efforts of data scientists can serve coercive functions if they are not grounded in the needs and desires of the communities supposedly being served. In order to promote long-term structural change and social justice, larger shifts in data science practice are necessary.

## 3.4 Stage 4: Practice

The final stage is to develop new modes for what it means to practice data science. Achieving changes along these lines requires developing new epistemologies, methodologies, and cultures for data science. While the path ahead remains somewhat speculative, several broad directions are clear.

### 3.4.1 Participatory data science

Data scientists must abandon their desire for a removed objectivity in favor of participation and deliberation among diverse perspectives. STS scholar Donna

Haraway argues for a new approach centered on "situated knowledges": she articulates the need "for a doctrine and practice of objectivity that privileges contestation and deconstruction", one that recognizes that every claim emerges from the perspective of a particular person or group of people[41]. Following this logic, the "neutral" data scientist who attempts to minimize position-taking must be replaced by a data science of situated values—a "participatory counterculture of data science"[82]. This perspective highlights the importance of groups such as Black in AI,⑯ LatinX in AI,⑰ Queer in AI,⑱ and Women in Machine Learning,⑲ all of which work to increase the presence of underrepresented groups in the field of artificial intelligence. Given that data science is influenced by practitioners' perceptions of problems and of how to address them, it is essential to encourage greater diversity in data science[83].

Complementing this participatory approach is for data science to focus more directly on "designing with" rather than "designing for" affected communities and social movements. Data scientists must develop procedures for incorporating a multitude of public voices into their work. When engineers privilege their own perspectives and fail to consider the multiplicity of needs and values across society, they tend to erase and subjugate those who are already marginalized[84−90]. To avoid participating in these oppressive (even if inadvertent) acts, data scientists must center affected communities in their work. One approach toward this end is the principle of "Nothing about us without us", which has been invoked in numerous social movements (in particular, among disability rights activists in the 1990s) to signify that no policies should be developed without direct participation from the people most directly affected by those policies[91]. The Design Justice Network articulates a powerful enactment of these values, with its commitments to "center the voices of those who are directly impacted" and to "look for what is already working at the community level"[92].

This type of approach represents a notable departure from traditional data science practice and values—efficiency and convenience—toward democracy and empowerment. A great deal of work in recent years has exemplified this approach[79, 93−100]. Mechanisms for participatory design and decision making—such as charrettes, participatory budgeting, and co-production—present further models of designing with communities. Any participatory practices should entail not just the design of an algorithm, but also broader questions such as whether an algorithm should be developed in the first place and how it should be used. Additionally, an essential component of developing a more democratic data science is to bring data scientists, technology companies, and governments within the ambit of democratic oversight and accountability[101].

**3.4.2 New methods and cultures**

Adapting data science to a political orientation and to participatory practices will require new methods. Broadly speaking, data science must move toward a "critical technical practice" that rejects "the false precision of mathematical formalism" to engage with the political world in its full complexity and ambiguity[102]. It is necessary to expand the bounds of algorithmic reasoning, shifting from the dominant method of "algorithmic formalism" to the alternative method of "algorithmic realism" that better accounts for the realities of social life and the impacts of algorithmic interventions[68].

As a central component of this evolution, the field should change its internal structures to incentivize greater attention to the implementation and impacts of data science. To embrace justice and tackle the most pressing social issues related to algorithms, data science must take a more expansive approach to research contributions that looks for more than technical contributions. Actually improving people's lives with data science requires far more than just developing a technical tool—it also requires thoughtfully adapting data science methods to the needs of a particular organization or community[37]. If data scientists are to contribute to improving society, they need a more rigorous methodology for ensuring that data science tools produce beneficial impacts when implemented in real-world contexts. New workshops, conferences, and journals will be essential mechanisms for fostering novel methods that blend technical and nontechnical approaches.

Along these lines, data scientists must also adopt a reflexive political standpoint that grounds their efforts in

⑯ https: //blackinai.github.io/

⑰ http://www.latinxinai.org/

⑱ https: //sites.google.com/view/queer-in-ai/

⑲ https: //wimlworkshop.org

rigorous evaluations of downstream social and political consequences. What ultimately matters is not how an algorithm performs in the abstract, but what impacts an algorithm has when introduced into complex sociopolitical environments. Data scientists cannot be expected to perfectly predict the impacts of their work—the entanglements between technology and society are far too complex. However, through collaborations with communities and with scholars from other fields, well-grounded analyses are possible. Just as data scientists would demand rigor in claims that one algorithm is superior to another, they should also demand rigor in claims that a technology will have any particular impacts. Toward this end, one necessary direction for future research is to develop interdisciplinary frameworks that will help data scientists consider the downstream impacts of their interventions. This requires being mindful of the various forms of "indeterminacy" that may lead an algorithm to generate different impacts than its developers expect[68].

As one example of a reform that emphasizes impacts as a central concern, in 2018 the ACM Future of Computing Academy proposed that peer reviewers should consider the potential negative implications of submitted work and that conducting "anti-social research" should factor negatively into promotion and tenure cases[103]. Just two years later, the Neural Information Processing Systems Conference (NeurIPS)—one of the world's top AI conferences—announced that every paper at the 2020 conference must include a "broader impact" section that discusses the positive and negative social consequences of the research[104].

#### 3.4.3 Engaging with the broader political context

Of course, shifts in data science practice do not occur in a vacuum. Shifts in data science practice require broader structural reforms that contribute to a more just society. As historian Elizabeth Fee notes, "we can expect a sexist society to develop a sexist science; equally, we can expect a feminist society to develop a feminist science"[105]. Similarly, we can expect a militarized society of economic inequality to produce a militarized and unequal data science[106, 107].

Data scientists committed to social justice must work toward more structural reforms against the harms of digital technologies. For instance, building solidarity and power among workers can shift the development of data science away from the most harmful applications. In recent years, tech workers have organized against their companies' partnerships with the United States Departments of Defense and Homeland Security. Rather than perceiving themselves as "just an engineer", these technologists recognize their position within larger sociotechnical systems, recognize the connection between their work and its social ramifications, and hold themselves (and their companies) accountable for these impacts. Building on this movement, thousands of computer science students from more than a dozen US universities pledged in 2019 that they will not work for Palantir due to its partnerships with Immigration and Customs Enforcement (ICE)[108]. Data scientists should also provide support for communities and activists organizing in opposition to oppressive algorithms.

Data scientists alone cannot be held responsible for promoting social and political progress. They are just one set of actors among many. The task of data scientists is not to eradicate social challenges on their own, but to act as thoughtful and productive partners in broad coalitions and social movements striving for a more just society.

## 4 Conclusion

The field of data science must abandon its self-conception of being neutral to recognize how, despite not being engaged in what is typically seen as political activity, data science logics, methods, and technologies shape society. Restructuring the values and practices of data science around a political vision of social justice will not be easy or immediate, but it is necessary. Given the political stakes of algorithms, it is not enough to have good intentions—data scientists must ground their efforts in clear political commitments and rigorous evaluations of the consequences.

As a form of political action, data science can no longer be separated from broader analyses of social structures, public policies, and social movements. Instead, the field must debate what impacts are desirable and how to promote those outcomes—thus prompting rigorous evaluations of the issues at hand and openness to the possibility of non-technological alternatives. Such deliberation needs to occur not just among data scientists, but also with scholars from other fields, policymakers, and communities affected by data science systems.

Recognizing data science as a form of political action will empower and enlighten data scientists with new frameworks to improve society. By deliberating about political goals and strategies and by developing new methods and norms, data scientists can more rigorously contribute to social justice.

## Acknowledgment

B. Green is grateful to the Berkman Klein Center Ethical Tech Working Group for fostering his thinking on matters of technology, ethics, and politics. B. Green also thanks Catherine D'Ignazio, Anna Lauren Hoffmann, Lily Hu, Momin Malik, Dan McQuillan, Luke Stark, Salomé Viljoen, and the reviewers for providing helpful discussions and suggestions.

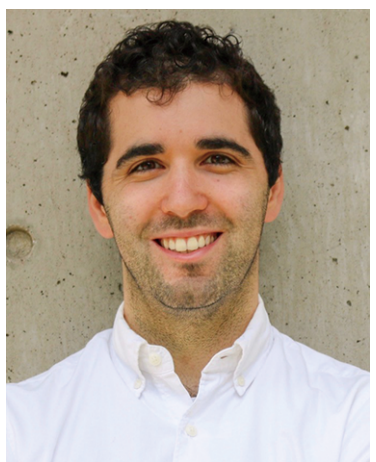

**Ben Green** is a postdoctoral scholar in the Society of Fellows and an assistant professor in the Gerald R. Ford School of Public Policy, University of Michigan . He received the PhD degree in applied math (with a secondary field in STS) from Harvard University and the BS degree in mathematics & physics from Yale College in 2020 and 2014, respactively.